\documentclass{article}

\usepackage{caption}
\usepackage{subcaption}
\usepackage{url}

\usepackage{graphicx}

\begin{document}

%%
%% The "title" command has an optional parameter,
%% allowing the author to define a "short title" to be used in page headers.
\title{INTERACT: An authoring tool that facilitates the creation of human centric interaction with 3d objects in virtual reality%%
%% The acknowledgments section is defined using the "acks" environment
%% (and NOT an unnumbered section). This ensures the proper
%% identification of the section in the article metadata, and the
%% consistent spelling of the heading.
\thanks{This paper was supported by the European Union’s Horizon HADEA research and innovation program under grant Agreement 101092851 XR2LEARN project.}
}

%%
%% The "author" command and its associated commands are used to define
%% the authors and their affiliations.
%% Of note is the shared affiliation of the first two authors, and the
%% "authornote" and "authornotemark" commands
%% used to denote shared contribution to the research.
\author{Rama Krishnan Gopal Ramasamy Thandapani\footnote{LS Group, Paris, France}\\
\texttt{rama.gopal@ls-group.fr}
\and
Benjamin Capel\footnotemark[2]\\
\texttt{benjamin.capel@ls-group.fr}
\and
Antoine Lasnier\footnotemark[2]\\
\texttt{antoine.lasnier@ls-group.fr}
%\affiliation{%
%  \institution{LS Group}
%  \city{Paris}
%  \country{France}
%}
\and
Ioannis Chatzigiannakis\footnote{Sapienza University of Rome, Rome, Italy}\\
\texttt{ichatz@diag.uniroma1.it}
%\orcid{0000-0001-8955-9270}
%\affiliation{%
%  \institution{Sapienza University of Rome}
%  \city{Rome}
%  \country{Italy}}
}

%%
%% By default, the full list of authors will be used in the page
%% headers. Often, this list is too long, and will overlap
%% other information printed in the page headers. This command allows
%% the author to define a more concise list
%% of authors' names for this purpose.
%\renewcommand{\shortauthors}{R.K.G.Ramasamy Thandapani, B.Capel, A.Lasnier, I.Chatzigiannakis}

\maketitle

%%
%% The abstract is a short summary of the work to be presented in the
%% article.
\begin{abstract}
A widespread adoption of Virtual, Augmented, and Mixed Reality (VR/AR/MR), collectively referred to as Extended Reality (XR), has become a tangible possibility to revolutionize educational and training scenarios by offering immersive, interactive experiences. In this paper we present \textsf{INTERACT}, an authoring tool for creating advanced 3D physics-based Intelligent Tutoring Systems (ITS) by individual developers or small-scale development teams. \textsf{INTERACT} is based on a cutting edge physics engine allowing realistic interactions such as collision detection and ergonomic evaluations. We demonstrate the benefits of \textsf{INTERACT} by developing a set of training scenarios for a use case of a Laser cutting machine. The use case illustrates the numerous possibilities such as creating interaction with objects, ease of configuring a scenario and how to design the visual effects to the machine.
\end{abstract}

%%
%% The code below is generated by the tool at http://dl.acm.org/ccs.cfm.
%% Please copy and paste the code instead of the example below.
%%
%\begin{CCSXML}
%<ccs2012>
%   <concept>
%       <concept_id>10003120.10003145.10003151.10011771</concept_id>
%       <concept_desc>Human-centered computing~Visualization toolkits</concept_desc>
%       <concept_significance>300</concept_significance>
%       </concept>
%   <concept>
%       <concept_id>10010405.10010489.10010491</concept_id>
%       <concept_desc>Applied computing~Interactive learning environments</concept_desc>
%       <concept_significance>500</concept_significance>
%       </concept>
%   <concept>
%       <concept_id>10010405.10010489.10010490</concept_id>
%       <concept_desc>Applied computing~Computer-assisted instruction</concept_desc>
%       <concept_significance>500</concept_significance>
%       </concept>
% </ccs2012>
%\end{CCSXML}
%
%\ccsdesc[300]{Human-centered computing~Visualization toolkits}
%\ccsdesc[500]{Applied computing~Interactive learning environments}
%\ccsdesc[500]{Applied computing~Computer-assisted instruction}

%%
%% Keywords. The author(s) should pick words that accurately describe
%% the work being presented. Separate the keywords with commas.
%\keywords{Authoring Tools, Intelligent Tutoring Systems (ITS), Human-centric interaction, Virtual Reality, Unity}

%\received{26 May 2023}
%\received[revised]{12 March 2009}
%\received[accepted]{5 June 2009}

%%
%% This command processes the author and affiliation and title
%% information and builds the first part of the formatted document.

\section{Introduction}

Currently, there is a significant surge in interest surrounding Virtual, Augmented, and Mixed Reality (VR/AR/MR), collectively referred to as Extended Reality (XR)~\cite{FASTBERGLUND201831,chatzigiannakis2011urban}. 
The widespread adoption of XR in various innovative fields has become a tangible possibility due to groundbreaking research and continuous advancements of computing power, particularly on mobile devices, as well as the global expansion of connectivity. With the widespread availability of powerful mobile devices and continuous wireless internet connectivity, XR applications can now operate in real-time, on a large scale, and utilize extensive, up-to-date data. Furthermore, the existence of robust software components that offer advanced functionality in accessible, free, and open-source formats allows individual developers or small-scale development teams to create XR software with reasonable expectations of business success. XR is no longer confined to research laboratories or exclusive to established companies with significant investments in specialized software, hardware, and expert teams. Instead, it can be pursued as a feasible endeavor.

Emerging XR technologies have the potential to be applied in various fields with meaningful and beneficial outcomes. Beyond gaming and entertainment, XR applications in the field of education and training lag significantly behind, despite a wealth of literature highlighting the advantages of XR technologies in educational settings~\cite{chatzigiannakis2011implementing}. These technologies have been shown to expand knowledge areas, provide active learning experiences instead of passive information consumption, enhance understanding of complex concepts and subjects, minimize distractions during studying, stimulate creativity among students, and improve learning efficiency, among other benefits. Unfortunately, these features are currently perceived as luxuries accessible only to those who can afford them. Nevertheless, the continuous need for lifelong learning and the demand to upskill or reskill large audiences with varying levels of IT literacy, diverse cultural and educational backgrounds, and different languages will eventually make the integration of XR technologies in education and training a necessity. For instance, the manufacturing industry is currently grappling with the challenge of training the global workforce on the emerging industry 4.0 and 5.0 technologies.

In order to cater to evolving learning scenarios and educational and training requirements, there is a growing need for fast-paced Intelligent Tutoring Systems (ITS) in modern educational technologies. Such ITS must enable the adaptation of interactive and virtual experience-based learning activities. It is therefore crucial to develop authoring tools to allow the fast-paced creation of ITS offering support for a wider range of features beyond basic 3D object manipulation and avatar navigation in virtual environments. They should address aspects such as ergonomics, advanced physics for object and trajectory tracking, and more. To achieve this, new authoring tools need to be developed that facilitate the design of three-dimensional spaces, specifically in constrained physical environments like construction or manufacturing settings. These enablers will enhance the efficiency and effectiveness of creating immersive 3D spaces that meet the desired objectives in educational and training contexts.

In this work we present an authoring tool delivered as a Unity plugin which we call \textsf{INTERACT} that can significantly reduce the time needed to develop an ITS. The plugin is a no-code generic tool for creating physics-based VR training scenarios. \textsf{INTERACT} is based on a cutting edge physics engine allowing realistic interactions such as collision detection and ergonomic evaluations. The plugin allows any of its users to create physically realistic VR simulations for multiple usages such as training for various fields of applications (heavy industry, education, energy) from 3D data (CAD or point clouds) that are to be imported in the authoring tool. We demonstrate how \textsf{INTERACT} can be used in practice by developing a training application for a laser cutting machine. The resulting XR application provides a virtual reality training scenario targetting the maintenance tasks of a machine. With the application, the user is taught how to use a laser cutting machine in a step by step process and also validated with a score in the end. The application also illustrates the numerous possibilities such as creating interaction with objects, ease of configuring a scenario and how to design the visual effects to the machine.

\section{Related Work}

Various XR-based authoring tools for developing ITS have been proposed in the relevant literature to study how they can be used to assist in knowledge creation~\cite{10.1145/3313831.3376722}. Initially authoring tools provided a low-level framework that required the author to provide code~\cite{10.1162/105474602317343640}. More recent tools are starting to follow a low-code approach that make high-fidelity ITS prototyping easier~\cite{10.1145/3379337.3415824}. Most of these authoring tools are generic and are used to develope proof-of-concept ITS with a main goal to introduce certain features and study how they are perceived by non-technical designers~\cite{10.1145/3379337.3415824},\cite{8699236}. 

Authoring tools encompass a variety of software products that offer functionalities for composing, editing, assembling, and managing multimedia objects. These tools have the potential to decrease obstacles and enhance the accessibility of ITS for both inexperienced and skilled users, as mentioned by Ososky~\cite{ososky2016practical}. Murray~\cite{murray2016coordinating} shares a similar perspective, asserting that apart from tools created for internal usage by extensively trained specialists, authoring tools, by their inherent nature, should be user-friendly for a predetermined target audience. During the past years a limited number of domain-specific authoring tools have been proposed for the creation of VR-based ITS~\cite{10108673}. For the case of industry 4.0 environments, ARAUM~\cite{erkoyuncu2017improving}, ARTA~\cite{gattullo2019towards} and WAAT~\cite{10.1007/978-3-030-58468-9_22} are among the very few authoring tools available. These tools focus on how to provide guidance of the industrial procedures that need to be performed by the user. 

In contrast to these approaches, the embedded physics engine allows \textsf{INTERACT} to provide realistic behaviour of the different components that comprise an assembly line and thus move beyong simply displaying the instructions. Different components such as the cables, frames and lenses can be properly handled within the XR environment thus providing an immersive learning experience. Moreover, since \textsf{INTERACT} is implemented as a plugin of Unity game engine, it can naturally incorporate elements of gamification which can reinforce the pedagogical principles of the learning scenarios. Different diffuclty levels, scoring systems and awards can help in maintain the user's interest and focus throughout the learning process.

\section{INTERACT}

\textsf{INTERACT} is an Unity plugin that allows users to create a human centric interaction with 3d objects in virtual reality. It helps the users with low technical knowledge to swiftly set up a 3D scene in UNITY, physicalize objects and gamify the scenario. The main features are summarized as follows:

\begin{itemize}

\item Embedded physics engine: This handles multi-body dynamics, collision detection, friction, and kinematics, providing realistic behavior for objects in a 3D environment. 

\item Advanced collision detection: This feature allows accurate and efficient detection of collisions between objects, even when dealing with complex models. It includes snapping of lenses, closing of a glass door on a steel frame

\item Cables: The software can simulate cables and flexible beams using finite element analysis, providing realistic representations of these elements in the 3D environment.This includes the laser arm belt that is simulated as the laser arm moves.

\item Grab: This feature allows users to directly manipulate 3D objects with their own hands, providing a more natural and intuitive way to interact with the virtual environment.Though the application is capable of having interaction from an accessory like Manus gloves, we decided to have the Grab interaction with the VR controller but the interaction in game will represent a human hand. 

\item Scenarization: This is a module designed for assembly training, which can be used to create and edit complex assembly scenarios. This is the module where it helps us to gamify the scenes and create a pedagogical scenario which is customised to Maintanence procedures in our application. 

\end{itemize}

\begin{figure}
	\centering
	\begin{subfigure}[b]{0.52\textwidth}
		\centering
		\includegraphics[width=\textwidth]{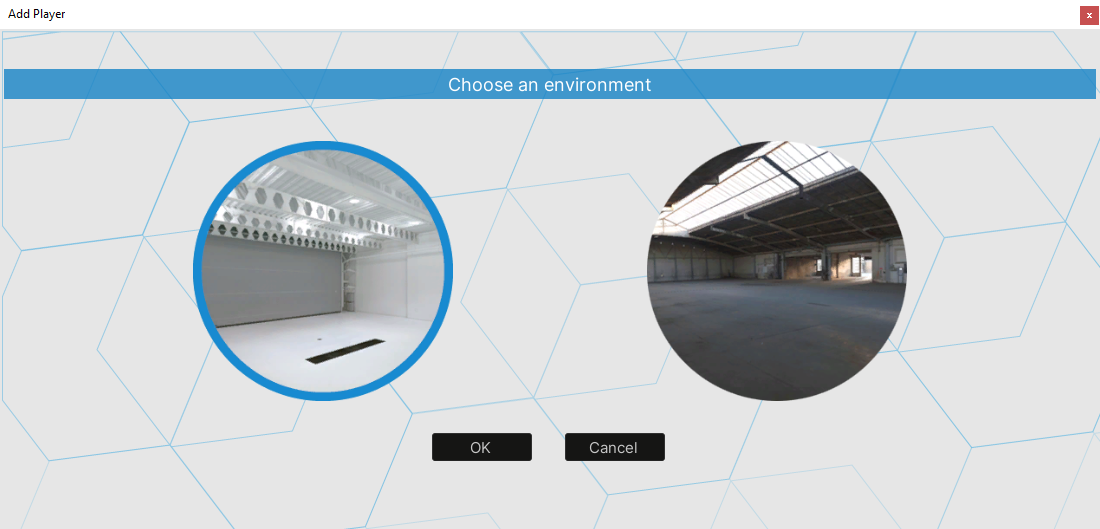}
		\caption{\label{fig:scene} Preconfigured environments}
	\end{subfigure}
	\hfill
	\begin{subfigure}[b]{0.43\textwidth}
		\centering
		\includegraphics[width=\textwidth]{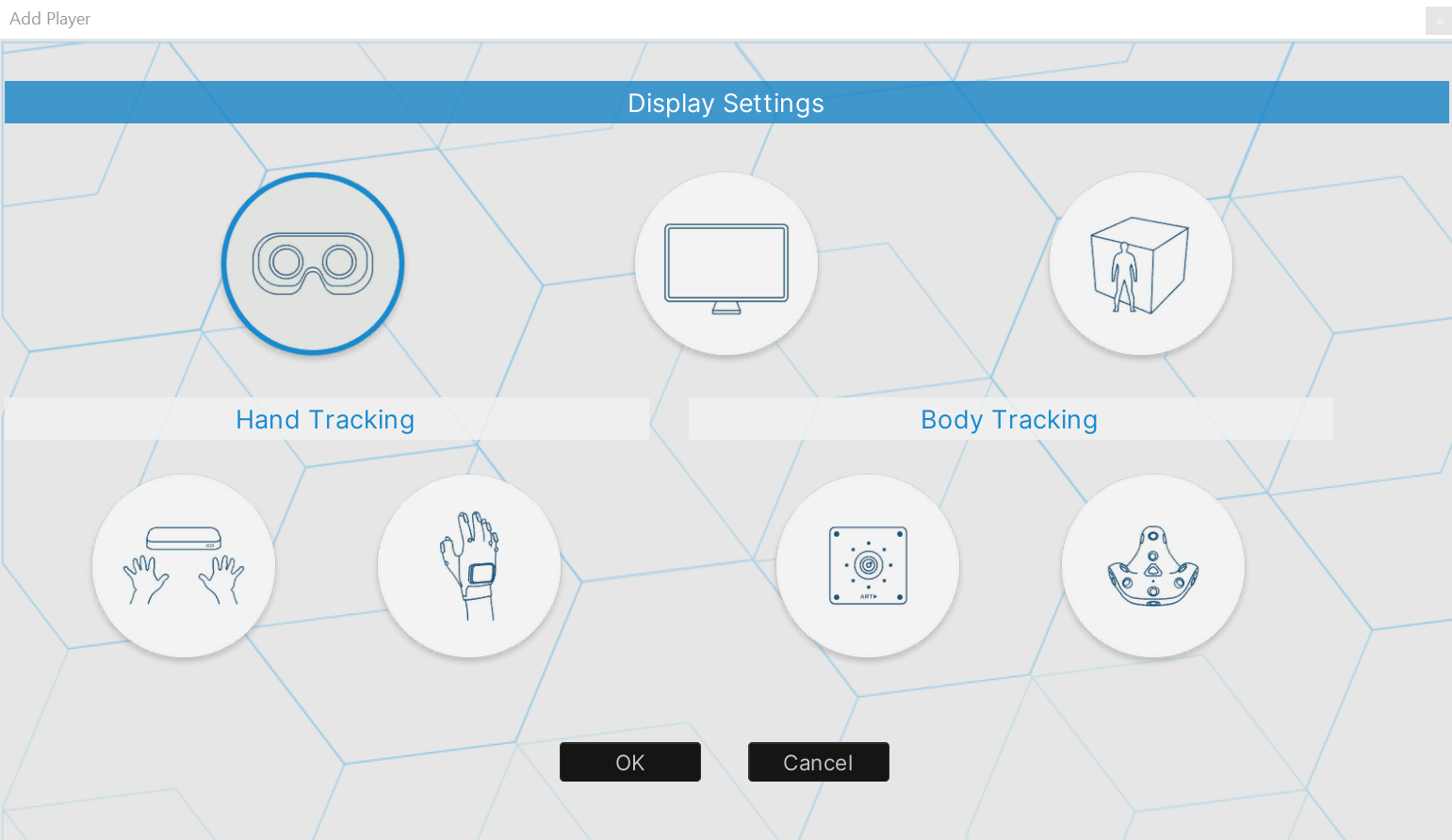}
		\caption{\label{fig:tools} Hardware devices}
	\end{subfigure}
	\caption{\label{fig:interact} Initial steps to create a new ITS}
\end{figure}

Initially the user needs to define among the two preconfigured environment: an empty white laboratory or a factory environment, as shown in Fig.~\ref{fig:scene}. The next step is to choose the hardware devices (Display Device, Hand Tracking, Body Tracking) through which the user will interact in XR environment, as shown in Fig.~\ref{fig:tools}.

The next step is the scenarization, where an \textit{assembly sequence} is provided. The assembly sequence is a critical part of the manufacturing process, as it ensures that the product is assembled correctly and efficiently.  Usually the assembly sequences constitute the practical exercises of a training application. The user is required to define the order in which different parts are put together to form a complete product. This typically involves a series of steps, in which each part is added to the product in a specific order. \textsf{INTERACT} helps to create such assembly sequences by visualizing the different parts and how they fit together. 

\begin{figure}
\centering
\includegraphics[width=.7\textwidth]{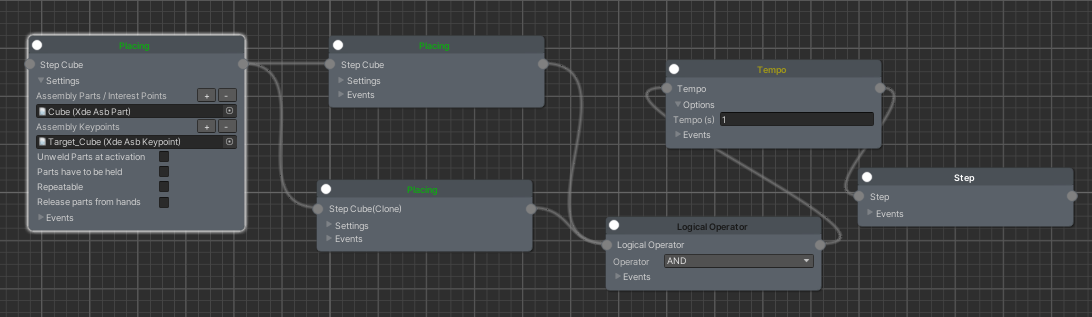}
\caption{\label{fig:scenarisation} The Scenario Graph used to create an assembly sequence comprised of multiple steps}
\end{figure}

In more details, \textsf{INTERACT} provides the \textit{Scenario Graph} to create a hierarchy of steps that create an assembly sequence. The user introduces 3D objects and indicates how they are connected through \textit{Placing Steps}. The user can encode rules to allow the learning scenario to unlock the next steps. For example, the assembly of a wheel starts only if the brake disk is in place AND the bolts have been properly screwed. Several options are available to describe your assembly process in the Scenario window, including time constraints that are required before proceeding to a subsequent step, or interaction with robots and actuators, etc. A scenario can also include \textit{Events}, that is actions that are  only triggered on specific conditions. For example to unweld or activate another part when a keypoint is reached.  
Fig.~\ref{fig:scenarisation} provides an example of the scenario graph and the series of steps that make up the assembly process.

The Scenario manager automatically handles the visual helpers in runtime (trajectories, ghost, instruction panel). In Simulation (when you switch to the PLAY mode), the transition between steps is done when the part to place reaches its target. For a detailed presentation of the implemented actions and events the interested reader is pointed to the dedicated website of \textsf{INTERACT}\footnote{\url{https://light-and-shadows.com/documentation/interact/scenarisation/\#welding-and-alignment}}.

\section{Laser Cutting Machine: a use case}

We use the \textsf{INTERACT} tool to develop an ITS on how to maintain a Trotec Speedy 400 laser cutting machine. The application aims to bringing together people and machines for safe, quick and easy access to trainings at a low cost, regardless of trainees’ location and hierarchical status. The resulting ITS provides a VR-based simulation for making the interaction with complex and dangerous machinery easy and safe. It illustrates the authoring process of immersive training applications using \textsf{INTERACT} from 3D data import to interaction configuration and scenario description. 

\begin{figure}
\centering
\includegraphics[width=.5\textwidth]{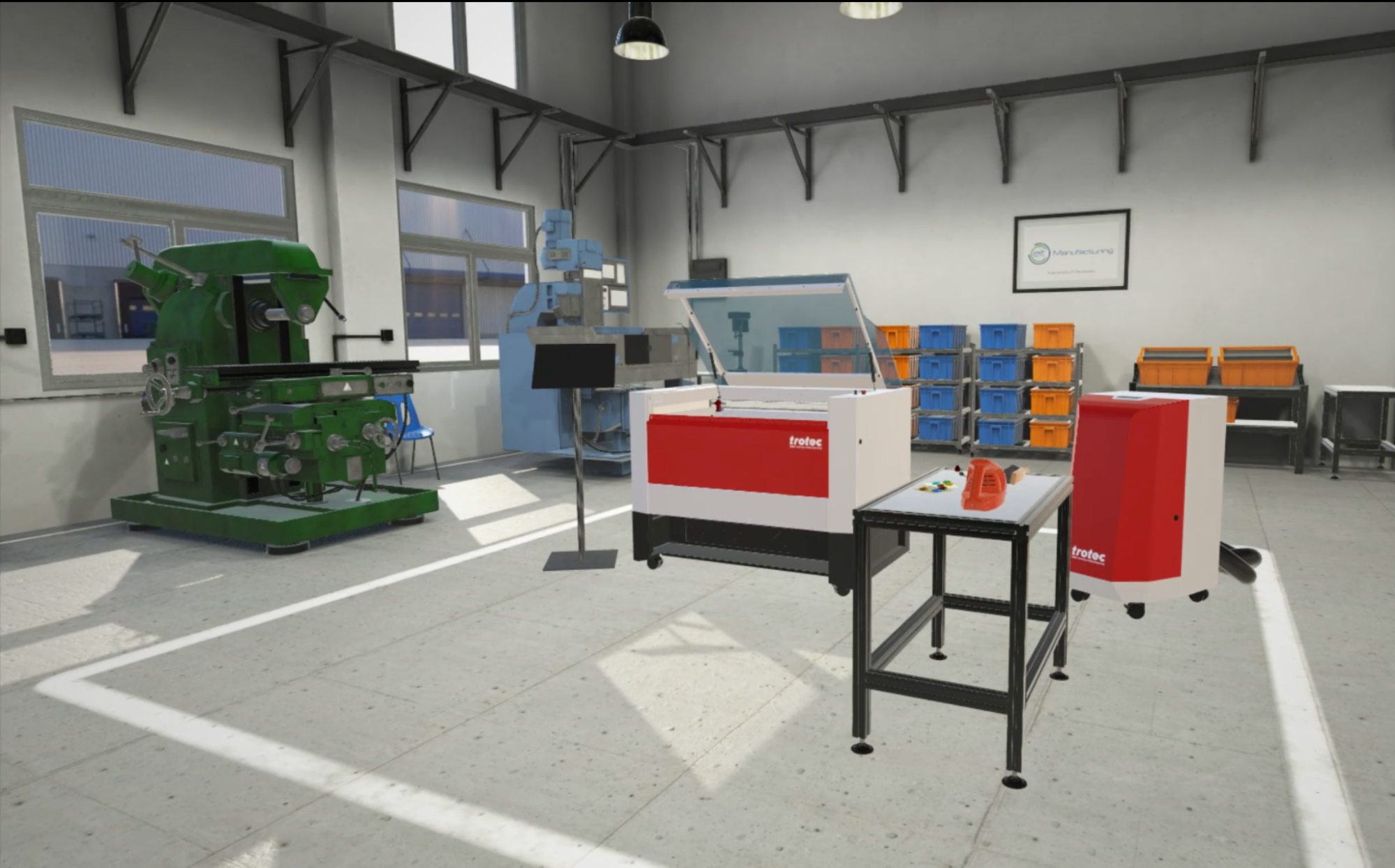}
\caption{\label{fig:laser1} Laboratory with the Trotec Speedy 400 laser cutting machine and a working table}
\end{figure}

\begin{figure}
	\centering
	\begin{subfigure}[b]{0.4\textwidth}
		\centering
		\includegraphics[width=\textwidth]{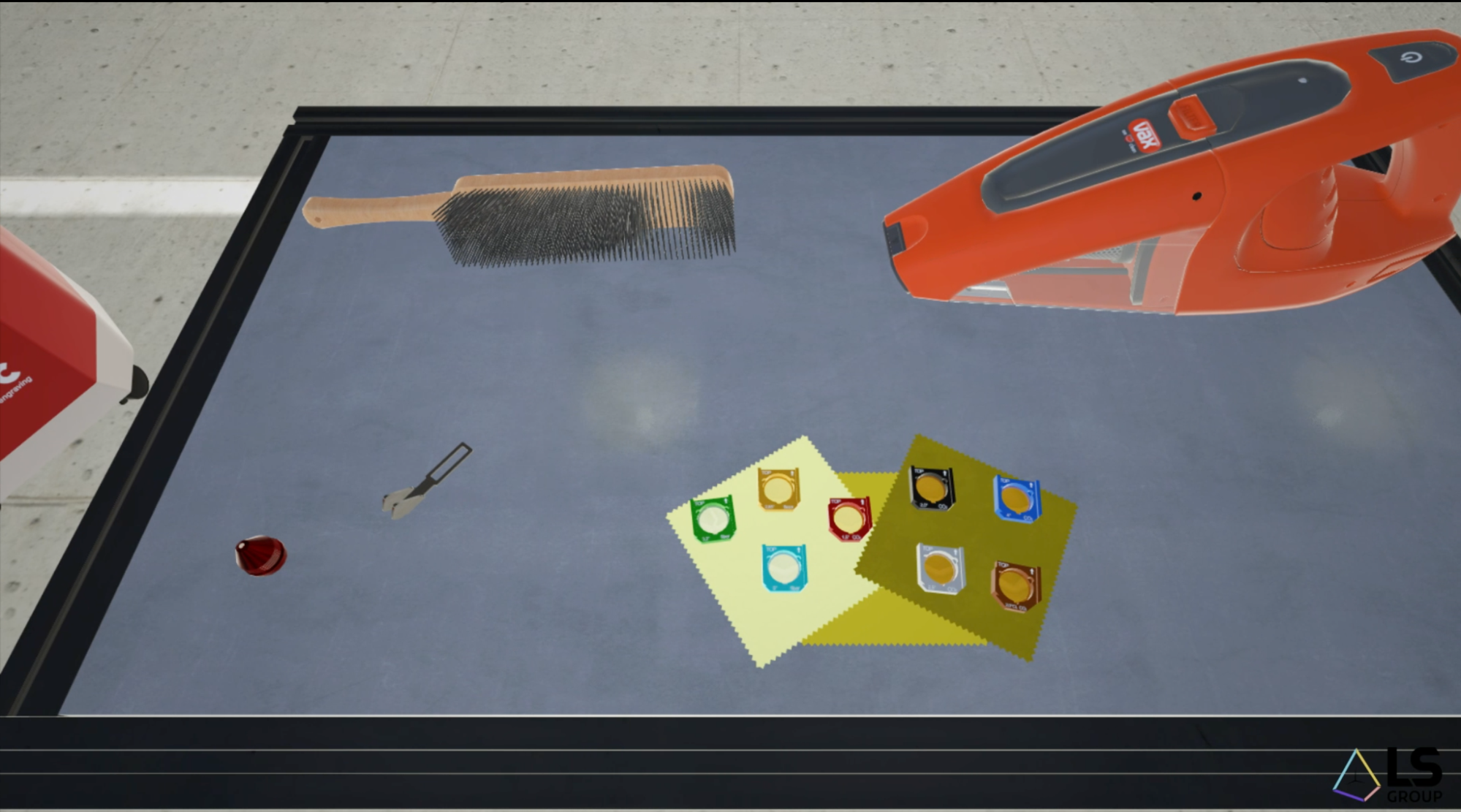}
		\caption{\label{fig:laser2a}}
	\end{subfigure}
	\hfill
	\begin{subfigure}[b]{0.57\textwidth}
		\centering
		\includegraphics[width=\textwidth]{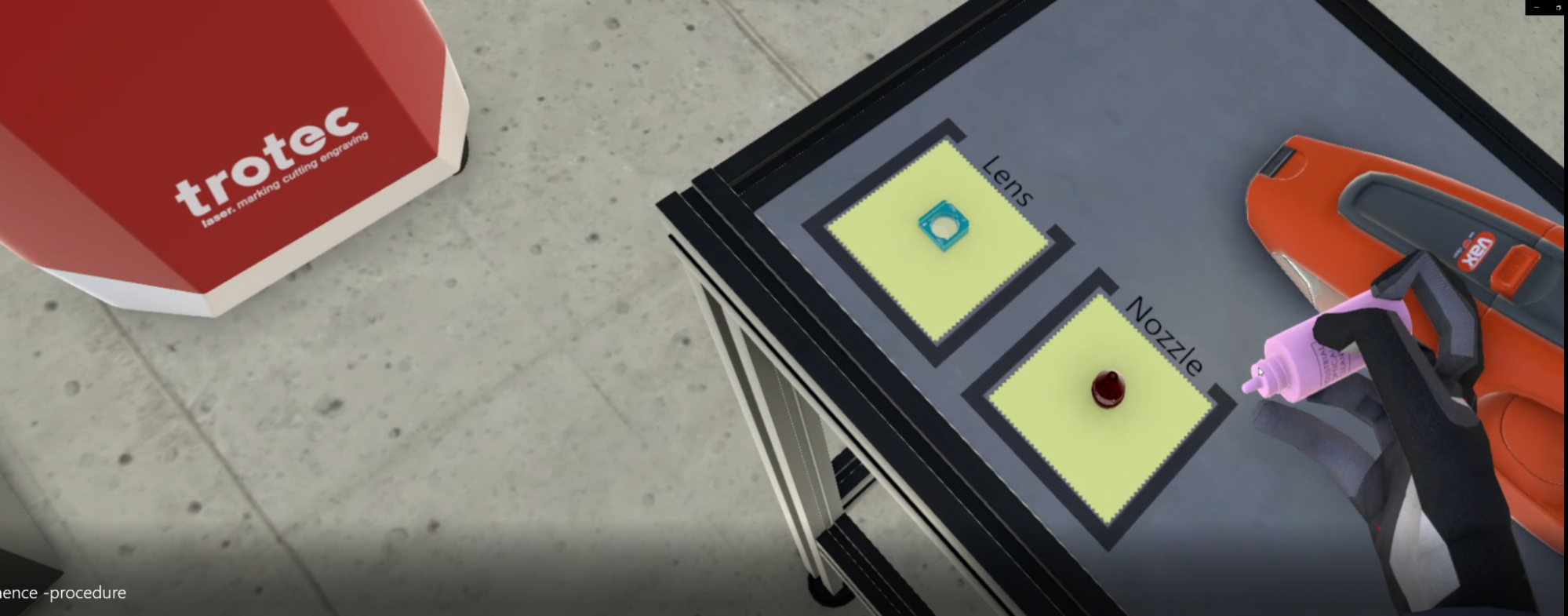}
		\caption{\label{fig:laser2b}}
	\end{subfigure}
	\caption{\label{fig:laser2} Various components of the laser cutting machine unmounted and cleaned using different tools}
\end{figure}

The application involves several steps to ensure the machine is properly maintained and continues to operate efficiently. Initially the user enters the scene where the Trotec Speedy 400 laser cutting machine is located along with a working table as shown in Fig.~\ref{fig:laser1}. The process typically involves turning off the machine and unmounting various components, such as the mirror, lens, and nozzle. The user approaches the machine and by following the instructions, turns off the machine and then unmounts the components indicated, as shown in Fig.~\ref{fig:laser2a}. In the sequel, the user is requested to wip the lens and nozzle with a fiber cloth as shown in Fig.~\ref{fig:laser2b}. The plate is wiped with a sponge, and the various components are remounted and the machine is turned back on. Removal or particles is also simulated in the application.  When the cleaning is complete, the components need to be remounted. Finally, the working table and main enclosure are then vacuumed to remove any dust or debris. 

A differenting aspect of this application is its commitment to user engagement and effective learning. The application incorporates elements of gamification, which are carefully aligned with pedagogical principles. The user is presented with multiple difficulty levels, which cater to different levels of experience and expertise. This strategy not only ensures that the content is appropriate for each user, but also provides a sense of progress and accomplishment as users advance through the levels. Furthermore, a scoring system is implemented at the end of each session. The scoring system is based on several factors, including time consumption, accuracy, and the use of hints or skip buttons. This comprehensive approach to scoring encourages users to optimize their performance while offering meaningful feedback.

\section{Conclusions and Future Work}

%Emerging technologies such as Extended Reality (XR) are revolutionizing educational and training scenarios, enhancing learning outcomes through immersive, interactive experiences. 
\textsf{INTERACT} provides a no-code approach to developing immersive ITS that is both accessible and safe. By simulating realistic industrial environments and processes, we are able to expose students and educators to experiences they might not otherwise have due to lack of real equipment or concerns about safety risks associated with operating industrial equipment. We demonstrate how \textsf{INTERACT} can be used in practice by developing a ITS that focuses on how to maintain a Trotec Speedy 400 laser cutting machine. Through this XR application, users can gain valuable practical knowledge and skills in maintaining a laser cutting machine – experiences that can be safely replicated and iterated, fostering deep and effective learning. 
%We are currently working on releasing the Laser Cutting Machine application and the \textsf{INTERACT} as open source projects so that developers/makers can use them for novel application development, prototyping and experimentation.

%%
%% Print the bibliography
%%
%\printbibliography
\bibliographystyle{plain}
\bibliography{software}

\end{document}